\documentclass[journal]{IEEEtran}
\usepackage{hyperref}
\usepackage{url}
\usepackage[cmex10]{amsmath}
\usepackage{amsfonts}

\usepackage{url}
\usepackage[numbers,sort]{natbib}
\usepackage[above,below]{placeins}

\usepackage[pdftex]{graphicx}
\usepackage{epstopdf}
\usepackage[subrefformat=parens,labelformat=parens,caption=false]{subfig}

\newcommand{\reals}{\mathbb{R}}
\newcommand{\err}[1]{\mathrm{err}_{\mathrm{#1}}}
\newcommand{\rank}[0]{\mathrm{rank}}

\title{Social Collaborative Retrieval}

\author{Ko-Jen~Hsiao,
        Alex~Kulesza,
        and~Alfred~O.~Hero,~\IEEEmembership{Fellow,~IEEE}
\thanks{This work was partially supported by ARO grant W911NF-12-1-0443. The paper is submitted to Journal of Selected Topics in Signal Processing for Social Networks for review on September 15, 2013. 

K.-J. Hsiao, A. Kulesza and A. Hero are with the Department of Electrical Engineering and Computer Science, University of Michigan, Ann Arbor, MI, 48109 USA (email: \{coolmark,kulesza,hero\}@umich.edu.)}}

\begin{document}

\maketitle

\begin{abstract}
Socially-based recommendation systems have recently attracted significant interest, and a number of studies have shown that social information can dramatically improve a system's predictions of user interests. Meanwhile, there are now many potential applications that involve aspects of both recommendation and information retrieval, and the task of \textit{collaborative retrieval}---a combination of these two traditional problems---has recently been introduced. Successful collaborative retrieval requires overcoming severe data sparsity, making additional sources of information, such as social graphs, particularly valuable. In this paper we propose a new model for collaborative retrieval, and show that our algorithm outperforms current state-of-the-art approaches by incorporating information from social networks. We also provide empirical analyses of the ways in which cultural interests propagate along a social graph using a real-world music dataset.

\end{abstract}

\section{Introduction}

\IEEEPARstart{C}{ollaborative} filtering (CF) and related recommendation techniques, which aim to automatically predict the interests of users and make personal recommendations of music, movies, products, or other items, have been both intensively studied by researchers and successfully deployed in industry during the past decade \cite{su2009survey, linden2003amazon, sarwar2001item, breese1998empirical}.  Recently, \citet{weston2012latent} proposed extending CF to a setting termed collaborative retrieval (CR), in which 
recommendations are made with respect to a particular \textit{query context}; for instance, a user might be looking for music in a particular genre or movies similar to a recent favorite.  In these situations, otherwise accurate recommendations will become irrelevant.  Similarly, a shopping website might want to deliver a list of recommended items to a user based on their browsing history.  In this case the recently viewed pages act as a sort of query, and we would like recommendations that are specific both to the query and to the user.  \citet{weston2012latent} proposed the first algorithm to solve CR problems, called latent collaborative retrieval (LCR).

However, several important issues remain.  While it is well-known that CF models often contend with data sparsity problems since a large number of user-item pairs must be judged for compatibility based on a relatively small dataset, CR models suffer even more severely from sparsity, since the range of possible queries multiplies the number of judgments to be made.  Viewed as a matrix completion problem, traditional CF requires filling out a user $\times$ item matrix, where each entry indicates the relevance of a specific item to a specific user.  In the same light, CR can be seen as a tensor completion problem, where the goal is to fill out a (much larger) query $\times$ user $\times$ item tensor.  Techniques like singular value decomposition and non-negative matrix factorization (NMF) \cite{seung2001algorithms}, applied widely in CF, have recently begun to be extended to tensor models of this type \cite{karatzoglou2010multiverse, xiong2010temporal, zheng2010collaborative, symeonidis2008tag}; however, these methods typically do not accommodate the ranking losses used for CR, and sparsity remains a major concern.  In this work, we propose to deal with data sparsity in CR by incorporating an additional (but often readily available) source of information: social networks.

In recent years social networking sites have become extremely popular, producing significant insights into the ways in which people connect themselves and interact.  Information derived from these networks can be used to help address the sparsity problem faced by recommender systems, for instance by propagating information from a user's social connections to fill in gaps in the recommendation matrix.  A variety of CF models augmented with social information have recently been proposed
\cite{ma2008sorec, purushotham2012collaborative, ma2011recommender, carmel2009personalized, konstas2009social}; these include state-of-the-art methods like Collaborative Topic Regression with social matrix factorization \cite{purushotham2012collaborative}, which is based on LDA \cite{blei2003latent}, and Probabilistic Matrix Factorization (PMF) \cite{salakhutdinov2008probabilistic,ma2008sorec}.  There has also been interest in so-called \textit{trust-aware} recommendation methods \cite{bedi2007trust, ma2009learning, massa2004trust, massa2007trust, o2005trust}, which are similar in spirit but inherently limited compared with using real social networks \cite{ma2011recommender}.  However, social information has not yet been employed for collaborative retrieval, which arguably stands to benefit even more due to data sparsity.  In this paper we set out to fill this gap.

We propose an approach we call social collaborative retrieval (SCR), building on the latent collaborative retrieval (LCR) model of \citet{weston2012latent} by integrating social networking data.  As in LCR, our algorithm sets out to optimize the top-ranked items retrieved for a given user and query, but we incorporate a regularization penalty that encourages a low dimensional embedding of each user to be similar to those of the user's nearest social neighbors.  On a collaborative retrieval task using real-world artist ratings from the Last.fm music dataset, our proposed algorithm significantly outperforms LCR, as well as baseline CF methods based on non-negative matrix factorization and singular value decomposition, particularly when a smaller training set leads to increased data sparsity.

The rest of this paper is organized as follows. In Section \ref{SCR} we provide a brief overview of latent collaborative retrieval (LCR) \citep{weston2012latent}, and then describe our proposed SCR algorithm in detail. Section \ref{SIA} contains an empirical analysis of the Last.fm social networking dataset, and finally we present experimental results evaluating the performance of SCR in Section \ref{exp}.

\section{Collaborative retrieval}
\label{SCR}

The goal of collaborative retrieval is to produce a ranked list of items that are of interest to a particular user given a particular query.  
While a natural approach to this problem might be to simply filter a set of unconstrained CF recommendations for the specified user using the query---or, conversely, to filter a set of generic search results for the query using the user's profile---these pipeline approaches fail to account for interactions between the query and the user.  For instance, two users might have very different interpretations of the query ``jazz'', despite having broadly similar preferences among artists.  The idea of CR is to obtain more informative results through a unified approach.

We therefore formalize the problem by defining a single score function $f(q,u,a)$ to represent the relevance of a given item $a$ with respect to both a query $q$ and a user $u$.  If we enumerate all users, queries, and items, we can think of this score function as specifying the values of a rating tensor ${\bf R} \in \reals^{|\mathcal{Q}|\times|\mathcal{U}|\times|\mathcal{A}|}$, where $\mathcal{Q}$ is the set of all queries, $\mathcal{U}$ is the set of users, and $\mathcal{A}$ is the set of items.  However, in practice we usually only care about the top $k$ items retrieved (for some small constant $k$) for a given user and query, and our evaluation metrics will share this property as well.  (We discuss specific error measures in Section~\ref{sec:errormetrics}.)  Thus, learning a score function that can correctly \textit{rank} items for a given user-query pair is more important than learning one which can correctly approximate the full set of entries in ${\bf R}$.  The objectives that we use to learn the parameters of the score function will therefore involve a measure of error on such top-$k$ ranked lists. 

We next briefly review the existing latent collaborative retrieval model for this problem, and then introduce our model using social information.  Finally, we discuss the optimization needed to learn the parameters of the model.

\subsection{Latent collaborative retrieval}

Latent collaborative retrieval (LCR) \cite{weston2012latent} was the first algorithm proposed to solve collaborative retrieval problems.  The central idea is to embed users, queries, and items in a common $n$-dimensional space in which they can be compared using linear operations.  ($n$ is a hyperparameter that is typically tuned on a validation set.)  Formally, LCR is parameterized by matrices $S \in \reals^{n\times |\mathcal{Q}|}$, $V \in \reals^{n\times|\mathcal{U}|}$, and $T \in \reals^{n\times |\mathcal{A}|}$, which give the low-dimensional representations of queries, users, and items, respectively.  Additionally, for each user $u$ a matrix $U_u \in\reals^{n\times n}$ encodes the user-specific relationship between queries and items.  The scoring function $f$ is then defined as
\begin{equation}
  \label{lcreq}
  f(q,u,a) = S_q^{\top}U_uT_a + V_u^{\top}T_a~,
\end{equation}
where $S_q$ is the column of $S$ corresponding to query $q$, $T_a$ is the column of $T$ corresponding to item $a$, and $V_u$ is the column of $V$ corresponding to user $u$.  Intuitively, the first term in Equation~(\ref{lcreq}) measures the similarity between the query and the item under a linear transformation that is dependent on the user.  The second term is independent of the query and can be viewed as a bias term which models user preferences for different items.  Since for a given instance of a CR task the query and user are held fixed, there is no need for the scoring function to include a term like $S_q^{\top}\cdot V_u$, which would measure the compatibility of a user and a query.  However, interactions between the user and the query that pertain to actual item recommendations can be expressed in the first term.  If there are significant non-user-specific aspects of the compatibility between queries and items (i.e., a $S_q^{\top}\cdot T_a$ term), these can simply be absorbed into the first term and need not appear separately.

The parameters of the LCR scoring function are learned by optimizing a chosen error metric over a training set; we discuss some such metrics and other details in Section~\ref{sec:errormetrics}.

To aid in generalization, and to avoid the potentially prohibitive enumeration of queries, Equation~(\ref{lcreq}) can be generalized using features.  In this case $\Phi_Q(q)$, $\Phi_U(u)$ and $\Phi_A(a)$ are vector-valued feature maps for queries, users, and items, respectively, and $S$, $T$, and $V$ are linear maps from feature vectors to the embedded $n$-dimensional space.  The feature-based scoring function is given by
\begin{equation}
  f(q,u,a) = \Phi_Q(q)^{\top}S^{\top}U_uT\Phi_A(a)
             + \Phi_U(u)^{\top}V^{\top}T\Phi_A(a)~.
\end{equation}
If the feature maps are simple characteristic vectors, with a unique feature for each query, user, and item, then we recover the simpler form of Equation~(\ref{lcreq}).  Features of this type can also be used for content-based recommendation models (see \cite{weston2012latent}).  For our purposes, we simply note that this feature-based formulation can be easily extended to SCR, but for simplicity we focus on models of the type shown in Equation~(\ref{lcreq}).

\subsection{Social collaborative retrieval}

In the real world, people often turn to their friends for recommendations of musics, movies, or products. Here, we apply this intuition to improve the performance of CR techniques on tasks where social information is available. Our approach, which we refer to as social collaborative retrieval (SCR), learns a scoring function using a combination of \textit{behavioral} and \textit{relational} error measures.  

Behavioral measures encourage the model to respect implicit similarities between users, items, and queries that are revealed by the training data.  For instance, the preferences of one user may be useful for recommending items to another user if the two users have expressed similar preferences in the past.  This is the traditional mode of operation for collaborative filtering, as well as for CR.

Relational measures, on the other hand, take account of explicitly labeled connections that (hopefully) reveal underlying similarities.  In this work, we employ a relational measure that encourages the scoring function to be smooth with respect to a social graph; that is, we assume that users who are social neighbors should, on average, have more similar preferences than those who are not.  (We validate this assumption empirically in Section~\ref{SIA}.)  The hope is that this relational measure term provides complementary guidance to the system when little is known about the behavior of a user.

For simplicity, and to make the fairest comparisons later, we use the same parameterization of the scoring function as LCR (Equation~(\ref{lcreq})); we have $n$-dimensional representations of queries, users, and items in matrices $S$, $V$, and $T$, respectively, as well as user-specific transformations $U_u \in \reals^{n\times n}$.  We additionally assume that a social graph $G$ is available, where $G(i,j) = 1$ whenever users $i$ and $j$ are linked, and $G(i,j) = 0$ otherwise.  We will sometimes refer to users who are linked as ``friends'', though of course the social graph may encode relations with varying semantics.  To bias the preferences of friends toward each other, we introduce a social error measure
\begin{equation}
  \label{socialerr}
  \err{social}(V,G) = \sum_{i,j,G(i,j)=1}\|\sigma(V_i^TV_j)-1\|^2~,
\end{equation}
where $\sigma(\cdot)$ is the sigmoid function
\begin{equation}
  \sigma(x) = \frac{1}{1+e^{-cx}}
\end{equation}
and $c$ is a hyperparameter.  This measure can be seen as a regularization penalty that is minimized when friends have identical, high-norm representations in $V$.  Notice that we do not penalize similarity among non-friends, since users may have similar tastes even though they are not friends.  Importantly, although we encourage friends to have similar representations $V_u$, we do not introduce such regularization for $U_u$ matrices, as this would tend to force friends to always receive the same results.  Intuitively, we expect that friends are likely to have similar taste in items, but we allow each their own particular querying ``style''.

Combining the relational measure in Equation~(\ref{socialerr}) with a behavioral measure $\err{behavior}$ that depends on the scoring function $f$ and the training set $X$ yields the SCR learning objective to be minimized:
\begin{equation}
  \label{screrr}
  \err{behavior}(f,X) + w_s \err{social}(V,G)~,
\end{equation}
where $w_s$ is a regularization hyperparameter.  In the following subsection we will discuss choices for $\err{behavior}$, as well as optimization techniques used to find the parameters in practice.

Similarity-based error measures related to Equation~(\ref{socialerr}) have been proposed by others, typically based not on a social graph but instead on measured similarities between users.  For example, the measured Pearson correlation of item ratings can be used as a similarity measure $Sim(i,j)$ between users $i$ and $j$, and this can be incorporated as in \cite{ma2011recommender}:
$
  \sum_{i=1}^{|\mathcal{U}|}\left\| V_ i - \frac{\sum_{j,G(i,j)=1}Sim(i,j)\times V_j}{\sum_{j,G(i,j)=1}Sim(i,j)}  \right\|^2~.
$
Through the paper we denote by $\|\cdot\|$ the L2-norm of a vector. However, accurately estimating similarities from data is often unreliable due to sparsity, especially in the CR setting.  Moreover, such measures make it difficult to easily recommend items to newer users; without a long history of ratings, we cannot know which established users they are similar to.  On the other hand, SCR requires an external source of information in the form of a social graph.  Social networks are increasingly ubiquitous, and, since they are by nature centralized, can often be reliable even when extensive training data for a specific CR task is not yet available.

SCR can be viewed as a blend of social networking, collaborative filtering, and information retrieval.  As a side benefit, in addition to providing improved recommendations for users under particular query contexts, SCR can potentially be used in the inverse to recommend new social links between users with similar preferences.  In this way SCR can strengthen the social network and improve its own predictions in the future. 

\subsection{Learning}
\label{sec:errormetrics}

The goal of SCR learning is to (efficiently) find parameters $S$, $V$, $T$, and $U_u$ that minimize the objective in Equation~(\ref{screrr}).
In this section we describe the formal learning setup, the specific behavioral measures used in our experiments, and the algorithm used to optimize the model parameters.

We assume we are given a training set $X$ containing $N$ training examples:
\begin{equation}
  X = \{({ q}_i , { u}_i ,{ a}_i,{ w}_i  )\}_{i = 1,2,...,N}~,
\end{equation}
where $q_i \in \mathcal{Q}$ is a query, $u_i \in \mathcal{U}$ is a user, $a_i \in \mathcal{A}$ is an item, and $w_i \in \reals_{>0}$ is a measure of relevance for the item ${ a}_i$ given the user ${ u}_i$ and the query ${ q}_i$.  
Importantly, we assume that the weights $w_i$ always have a positive connotation; that is, triples $(q,u,a)$ that do not appear in the training set implicitly have a weight of zero, and are therefore dispreferred to triples that do appear.  For instance, in our experiments, $w_i$ will be derived from the number of times a user listens to a particular musical artist.

The behavioral part of the objective, which measures the compatibility of the scoring function $f$ (defined by the model parameters) with the training set $X$, can take a variety of forms depending on the setting.  As noted earlier, we will focus on top-$k$ ranking losses that optimize the most important aspects of the model, rather than, say, filling out all entries of the tensor ${\bf R}$.

Following \citet{weston2012latent} we define the vector $\bar{f}(q,u)$, which contains predictions for all items in the database given query $q$ and user $u$. The $a^{th}$ entry of $\bar{f}(q,u)$, denoted $\bar{f}_a(q,u)$, is equal to $f(q,u,a)$.  

With this notation, we can define the \textit{Weighted Approximate-Rank Pairwise (WARP) Loss}, introduced in \cite{weston2010large}:
\begin{equation}
  \label{errwarp}
  \err{WARP}(f,X) = 
  \sum^{N}_{i=1}L\left(\rank_{{ a}_i}\left(\bar{f}( { q}_i,{ u}_i) \right) \right)~.
\end{equation}
Here $\rank_{a_i}\left(\bar{f}(   { q}_i,{ u}_i) \right)$ is the margin-based rank of item ${ a}_i$,
\begin{equation}
  \rank_{a_i}\left(\bar{f}(q_i,u_i)\right) = \sum_{b\neq a_i }\mathbb{I}[1+\bar{f}_b(q_i,u_i)\geq \bar{f}_{a_i}(q_i,u_i)]~,
\end{equation}
where $\mathbb{I}[\cdot]$ is the indicator function, and $L$ is a loss function: \begin{align}
  L(k) &= \sum_{i=1}^k \alpha_i\\
  \alpha_1\geq \alpha_2&\geq\alpha_3\geq\dots\geq0~,
\end{align}
with the values of $\alpha_r$ determining the additional penalty for each successive reduction in rank.  We choose $\alpha_r = 1/r$, which gives a smooth weighting over positions while assigning large weights to top positions and rapidly decaying weights to lower positions. 

Intuitively, the WARP loss prefers that the item $a_i$ is always ranked highest.  For each training example $i = 1,\dots,N$, the positive item $a_i$ is compared pairwise with all other (negative) items in the database.  If the score of another item is less than a margin of one from the score of $a_i$, this pair incurs a cost.  The WARP loss determines this cost based on the corresponding items' ranking positions and the choice of $\alpha$ parameters.  

We use the WARP loss in our experiments for comparison with prior work.  However, in our setting it ignores the relevance scores $w_i$ that are part of the training set; this can be inefficient, since the optimization cannot focus on the most important training examples.  We thus propose a modified behavioral measure that we refer to as the generalized WARP (gWARP) loss:
\begin{equation}
  \label{behaviorerr}
  \err{gWARP}(f,X) = \sum^{N}_{i=1}{ w}_iL\left(\rank_{{ a}_i}\left(\bar{f}({ q}_i,{ u}_i)\right)\right)~,
\end{equation}
where $\rank$ and $L$ are defined as before.  In our results, we refer to models learned under this loss as SCR-generalized, while models trained under the standard WARP loss are referred to simply as SCR.  We will derive the optimization procedure for the generalized WARP loss, since the standard WARP loss is a special case.

To minimize Equation~(\ref{screrr}), we employ stochastic gradient descent (SGD), choosing at each iteration a single training instance $i$ uniformly at random from the training set.  We then seek to optimize the objective for this single example; that is, we minimize
\begin{multline}
  { w}_i L\left( \rank_{{ a}_i}\left(\bar{f}({ q}_i,{ u}_i)\right) \right)\\
  + w_s \sum_{v,G({ u}_i,v)=1}\|1-\sigma(V_{{ u}_i}^{\top}V_v)\|^2~.
\end{multline}

Because it is expensive to compute the exact rank of an item $a_i$ when the total number of items is very large, the optimization procedure includes a sampling process at each step, as introduced in \citep{weston2010large}.  For the training sample $i$ chosen at the current iteration, negative items $b$ are sampled uniformly at random from $\mathcal{A}$ until a pairwise violation is found---that is, until $1+f({ q_i},{ u_i},b) > f(q_i,u_i,a_i)$.  If $K$ steps are required to find such a $b$, then the rank of $a_i$ can be approximated as
\begin{equation}
  rank_{a_i}(\bar{f}(q_i,u_i)) \approx  \left\lfloor\frac{|\mathcal{A}|-1}{K}\right\rfloor~,
 \label{rankapprox}
\end{equation}  
where $\lfloor \cdot \rfloor$ is the floor function. 

In each iteration of stochastic gradient descent, at most $|\mathcal{A}|-1$ sampling steps are required, since the right hand side of Equation~(\ref{rankapprox}) is constant (zero) for $K\geq |\mathcal{A}|-1$. Therefore at most $1+\text{min}\left(\frac{|\mathcal{A}|-1}{  rank_{a_i}(\bar{f}(q_i,u_i)) },|\mathcal{A}|-1\right)$ scores $f({ q_i},{ u_i},b)$ must be computed.  The worst case is when $a_i$ has rank one; however, in our experiments most items do not have small ranks, particularly during the early stages of training when the model still has large errors. As a result, rank approximation dramatically speeds up SGD in practice.  Note that SGD can also be parallelized to take advantage of multiple processors \citep{chu2007map,zinkevich2010parallelized}.

Following \citet{weston2010large}, the single-instance objective becomes
\begin{multline}
  { w}_i L\left(\left\lfloor \frac{|\mathcal{A}|-1}{K}\right\rfloor\right)\cdot|1-f({ q}_i,{ u}_i,{ a}_i)+f({ q}_i,{ u}_i,b) | \\
  +w_s \sum_{v,G({ u}_i,v)=1}\|1-\sigma(V_{{ u}_i}^{\top}V_v)\|^2~.
  \label{eqfuntimes}
\end{multline}
Rewriting Equation~(\ref{eqfuntimes}), we have
\begin{multline}
\label{obj}
C_i (1+(S_{{ q}_i}^{\top}U_{{ u}_i} + V_{{ u}_i}^{\top}       )(T_{b} - T_{{ a}_i}     ))\\
+w_s \sum_{v,G({ u}_i,v)=1}\|1-\sigma(V_{{ u}_i}^{\top}V_v)\|^2~,
\end{multline}
where $C_i = { w}_i\cdot L(\lfloor \frac{|\mathcal{A}|-1}{K}\rfloor)$. To speed up each gradient step, we only update the variables associated with the current violation pair; that is, we only update $S_{{ q}_i}$, $V_{{ u}_i}$, $T_{{ a}_i}$, $T_{b}$, and $U_{{ u}_i}$.  (In particular, we do not update the representations of ${ u}_i$'s friends $V_v$ for $G(u_i,v) = 1$.)  Now we can simply take the gradient of (\ref{obj}) to perform an update.  

The update for user ${ u}_i$'s low-dimensional embedding is
\begin{multline}
V_{{ u}_i} \leftarrow V_{{ u}_i} - \eta \left(C_i (T_{b} - T_{{ a}_i} )
\vphantom{\sum_1^2}
\right.\\
\left. + w_s\sum_{v,G({ u}_i,v)=1}(-2c\sigma(V_{{ u}_i}^TV_v)\left(1-\sigma(V_{{ u}_i}^TV_v))^2\right)\cdot V_j\right)~,
\end{multline}
or equivalently
\begin{equation}
V_{{ u}_i} \leftarrow V_{{ u}_i} - \eta C_i (T_{b} - T_{{ a}_i} ) + w_s'\sum_{v,G({u}_i,v)=1}b_v\cdot V_v~,
\end{equation}
where $b_v = 2c\sigma(V_{{ u}_i}^TV_v)(1-\sigma(V_{{ u}_i}^TV_v))^2 >0$, $w_s' = \eta w_s$, and $\eta$ is a learning rate parameter.  (Recall that $c$ is a hyperparameter for the sigmoid function.)  Thus at each gradient step, the user's low-dimensional embedding is updated toward the weighted mean of his or her friends' embeddings. 

Similarly, we have the following updates for the remaining parameters:
\begin{align}
S_{{ q}_i} &\leftarrow S_{{ q}_i} - \eta \left(\vphantom{\sum_1^2}C_i( U_{{ u}_i}(T_{b} -T_{{ a}_i}))\right)\\
T_{{ a}_i} &\leftarrow T_{{ a}_i} - \eta \left(\vphantom{\sum_1^2}C_i( -U_{{ u}_i}^T S_{{ q}_i} - V_{{ u}_i} )\right)\\
T_{b} &\leftarrow T_{b} - \eta \left(\vphantom{\sum_1^2}C_i( U_{{ u}_i}^T S_{{ q}_i} + V_{{ u}_i} )\right)\\
U_{{ u}_i} &\leftarrow U_{{ u}_i} - \eta \left(\vphantom{\sum_1^2}C_i\left( S_{{ q}_i} (T_{b} - T_{{ a}_i})^T\right)\right)~.
\end{align}

Finally, we constrain the parameters using
\begin{align} 
\|S_i\| &\leq L_S~,\quad i \in\{1,\dots,|Q|\}\\
\|T_i\| &\leq L_T~,\quad i \in\{1,\dots,|A|\}\\
\|V_i\| &\leq L_V~,\quad i \in\{1,\dots,|\mathcal{U}|\}
\end{align}
and project the parameters back on to the constraints at each step. These constraints can be viewed as additional regularizers that bound the lengths of vectors in $Q$, $A$, and $U$ with hyperparameters $L_S$, $L_T$, and $L_V$.

Once the SCR stochastic gradient training procedure converges, we take the learned parameters and apply them to predict scores for unseen test triples using Equation~(\ref{lcreq}).
\section{Social data analysis}
\label{SIA}

\begin{figure}[ht]
  \centering
  \subfloat[]{\includegraphics[width=6.8cm]{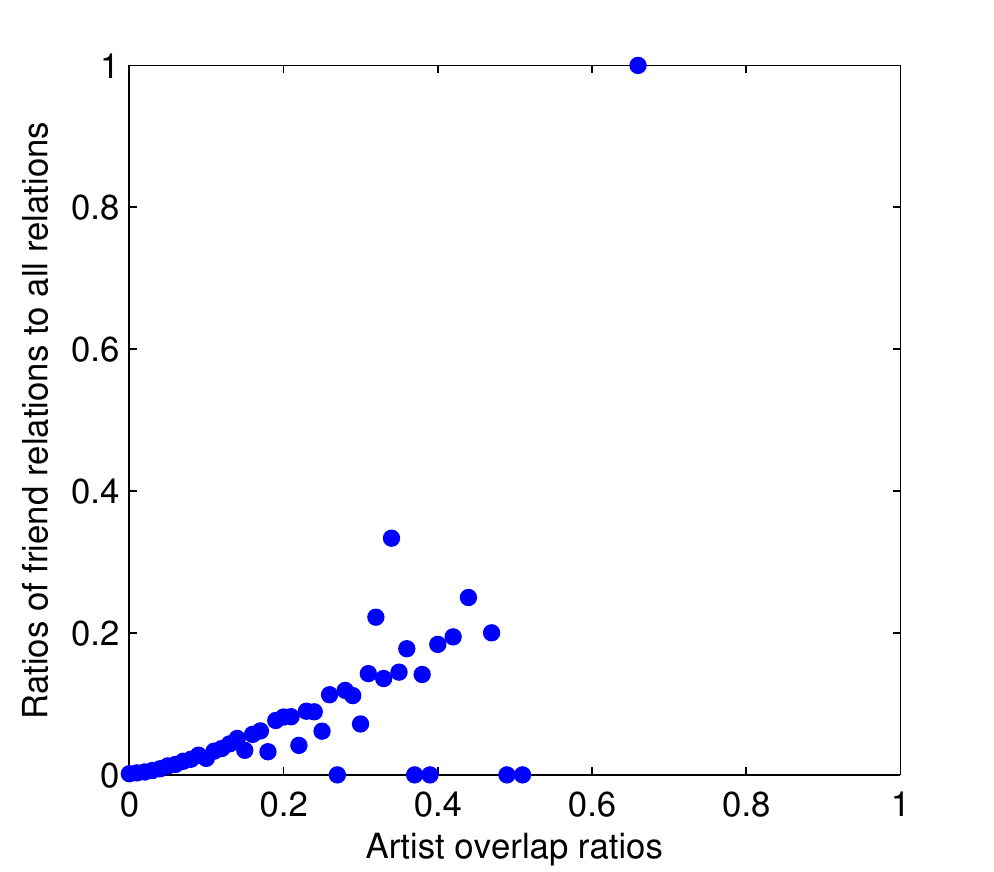}}\\
  \subfloat[]{\includegraphics[width=6.8cm]{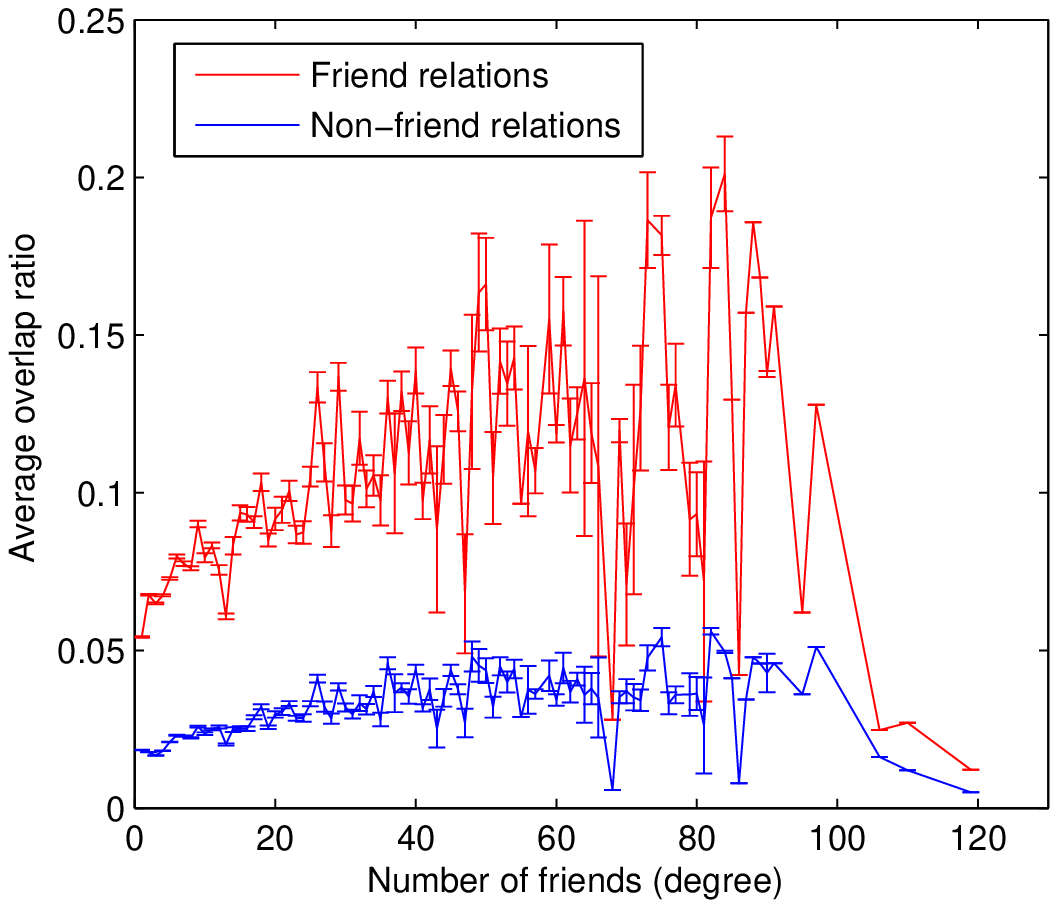}}
  \caption{(a) For user pairs having listened artist overlap ratios within the same interval, the proportion of friend relations among these pairs is shown. (b) Average artist overlap ratios among friends (red) and non-friends (blue) for users having different numbers of friends (degree).}
  \label{socialPlots}
\end{figure}

Before showing results that compare our proposed approach to existing
state-of-the-art methods, we first experimentally validate the fundamental assumption that friends, on average, have more in common than non-friends.

\subsection{Last.fm dataset}

In our experiments we use a real-word music dataset obtained from the Last.fm music website in May of 2011, \emph{hetrec2011-lastfm-2k} \cite{cantador2011second}. In this dataset each user is described by his or her listen counts for all musical artists in the database (items, in CR parlance), as well as any artist tags (queries) that might have been provided by each user.  

While the data contain more than ten thousand unique tags across all users, the vast majority of tags are used by only one user.  Typically these tags appear to be personal ``notes" rather than widely used genre distinctions.  To remove this noisy information, we throw out tags that are less frequent, keeping only the top 30 most common tags.  These tags were all used by at least 165 unique users, and generally correspond to genres of music; for example, the top 5 most popular tags are \emph{``rock''}, \emph{``pop''}, \emph{``alternative''}, \emph{``electronic''} and \emph{``indie''}. The Last.fm dataset contains listening histories for 1892 users and 17632 music artists.  A social graph is also included; on average each user has 13.44 friends.

\subsection{Shared musical interests}

Do friends share more preferences than non-friends?  This is a key question for our approach.  If the answer is no, it may not be useful to include social networks as a predictor variable in recommendation systems.  To estimate the similarity between two users' tastes for music, we compute the \emph{listened artists overlap ratio}, defined as
\begin{align}
  \mathrm{Sim}(i,j) = \frac{| A_i \cap A_j |}{| A_i \cup A_j |} \in [0,1] \quad\forall\ i,j~,
\end{align}
where $A_i$ is the set of artists listened to by user $i$.

We compute these overlap ratios for all $|\mathcal{U}| \choose 2$ user pairs. We then divide the range $[0, 1]$ of possible ratios evenly into 100 intervals, and calculate the fraction of the user pairs falling in each interval that are friends in the social graph.  Intuitively, we hope that users with greater similarity are more likely to be friends.
The result is shown in Figure \ref{socialPlots} (a). The percentage of realized \emph{friend} relations increases sharply as the artist overlap ratio increases. 

To reinforce this analysis, we also compute the average similarity between each user $i$ and his or her friends, as well as the average similarity between user $i$ and all other non-friend users, denoting the two numbers as $\beta^{i}_{friend}$ and $\beta^{i}_{non-friend}$.  Figure \ref{socialPlots} (b) shows the values of $\beta^{i}_{friend}$ and $\beta^{i}_{non-friend}$ averaged over users grouped by the number of friends they have in the social graph.  We can see that, regardless of how well-connected a user is, on average he or she has more in common with friends than with non-friends; moreover, the size of this effect increases for users with more friends.  Overall, these analyses support our assumptions regarding the use of social networks for recommendation and retrieval tasks on this dataset.

\FloatBarrier
\section{Experiments}
\label{exp}

\begin{figure}[t]
\begin{center}
\subfloat[]{
  \includegraphics[width=7 cm]{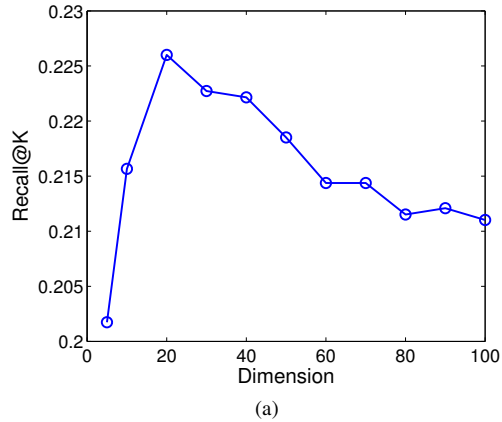}}\\
\subfloat[]{
  \includegraphics[width=7 cm]{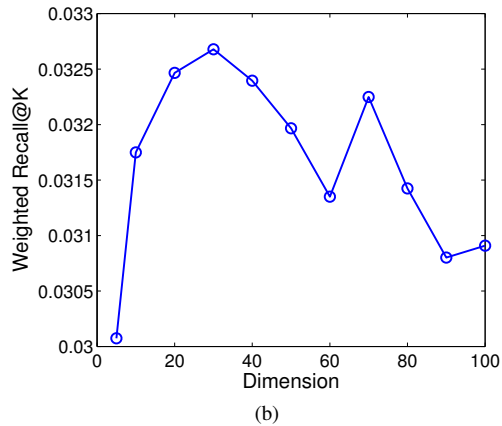}}
  \caption{(a) Recall@30 of SCR-generalized using different embedding dimensions.  (b) Weighted Recall@30 of SCR-generalized using different embedding dimensions.} 
  \label{recallDiffDimensionPlot}
  \end{center}
\end{figure}

\begin{figure}[t]
\begin{center}
\subfloat[]{
  \includegraphics[width=7.5 cm]{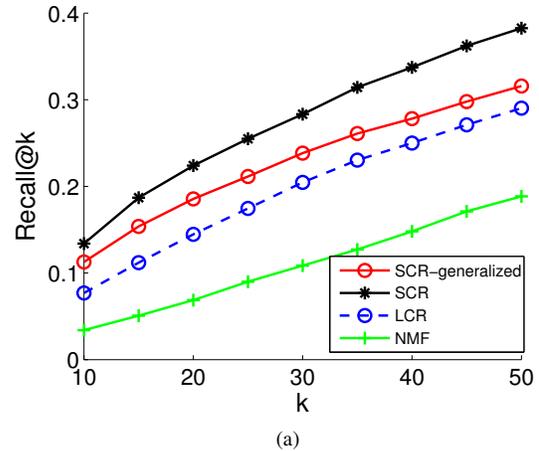}}\\
\subfloat[]{
  \includegraphics[width=7.5 cm]{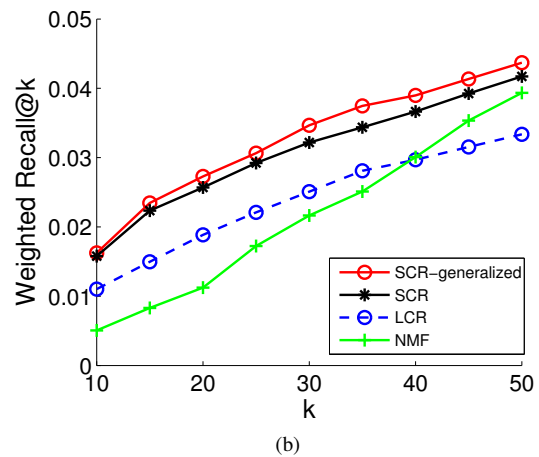}}
  \caption{(a) Recall at different values of k.  (b) Weighted recall at different values of k. } 
  \label{recallPlot}
  \end{center}
\end{figure}

\begin{figure}[t]
\begin{center}
\subfloat[]{
  \includegraphics[width=6.8cm]{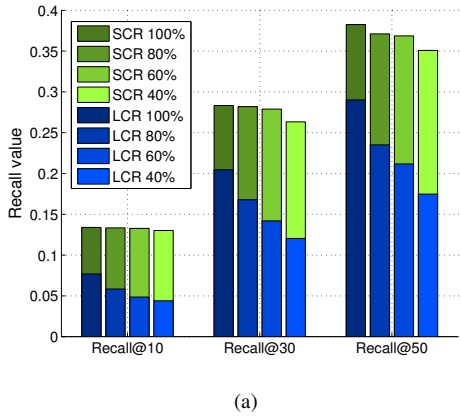}}\\
\subfloat[]{
  \includegraphics[width=6.8cm]{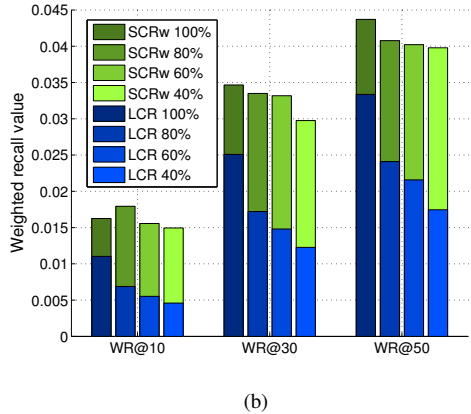}}
  \caption{(a) Recall for training data of different reduced sizes. (b) Weighted Recall for training data of different reduced sizes.} 
  \label{recallDiffTrainPlot}
  \end{center}
\end{figure}

\begin{figure}[t]
\begin{center}
\subfloat[]{
  \includegraphics[width=6.8cm]{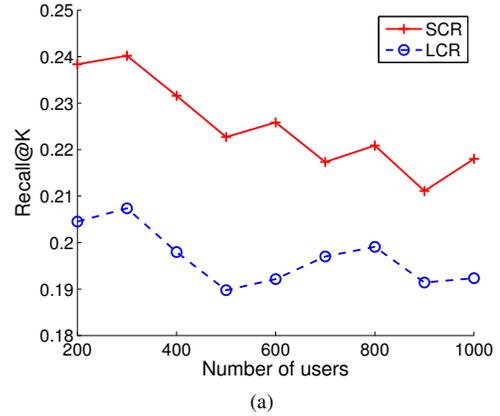}}\\
\subfloat[]{
  \includegraphics[width=6.8cm]{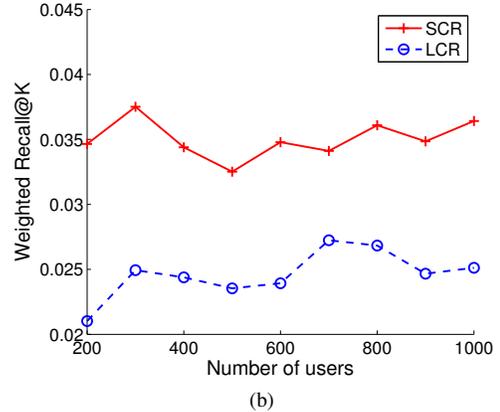}}
  \caption{(a) Recall@30 for datasets of different sizes.  (b) Weighted Recall@30 for datasets of different sizes.} 
  \label{recallDiffDatasetPlot}
  \end{center}
\end{figure}

\begin{figure}[t]
  \begin{center}
    \includegraphics[width=6.8 cm]{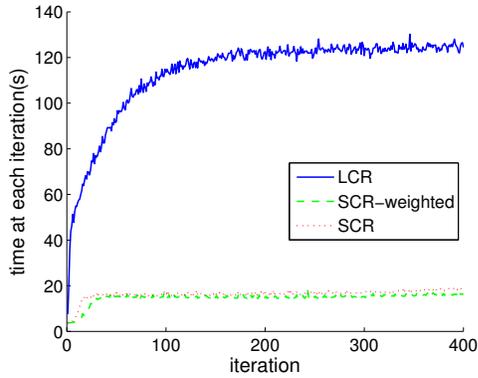}
    \caption{Runtime required for each training iteration of LCR and two versions of SCR on the Compact-\textit{lastfm-200} dataset.}
  \label{timeComparison}
  \end{center}
\end{figure}

\begin{figure}[t]
\begin{center}
\subfloat[]{
  \includegraphics[width=6.8cm]{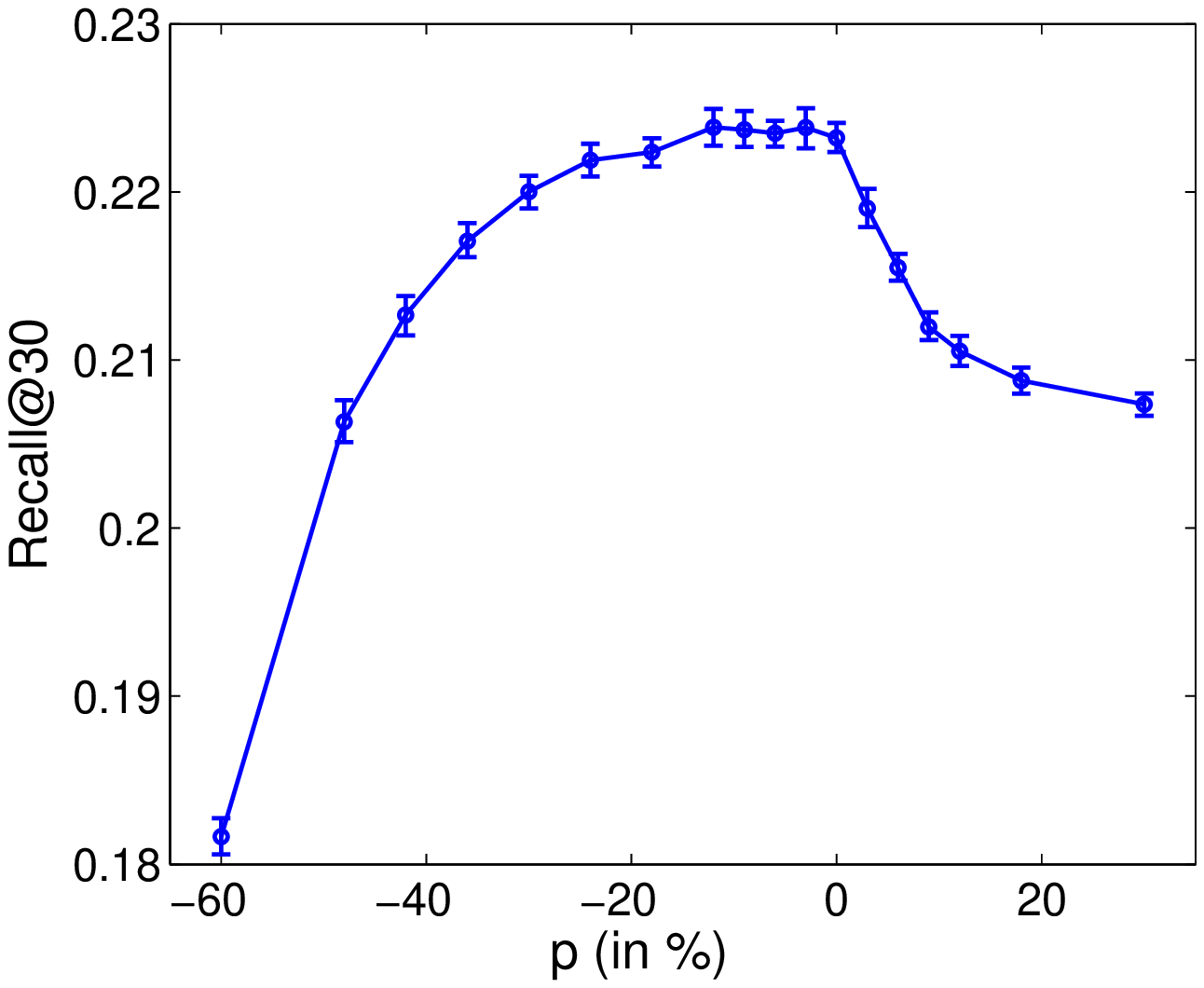}}\\
\subfloat[]{
  \includegraphics[width=6.8cm]
  {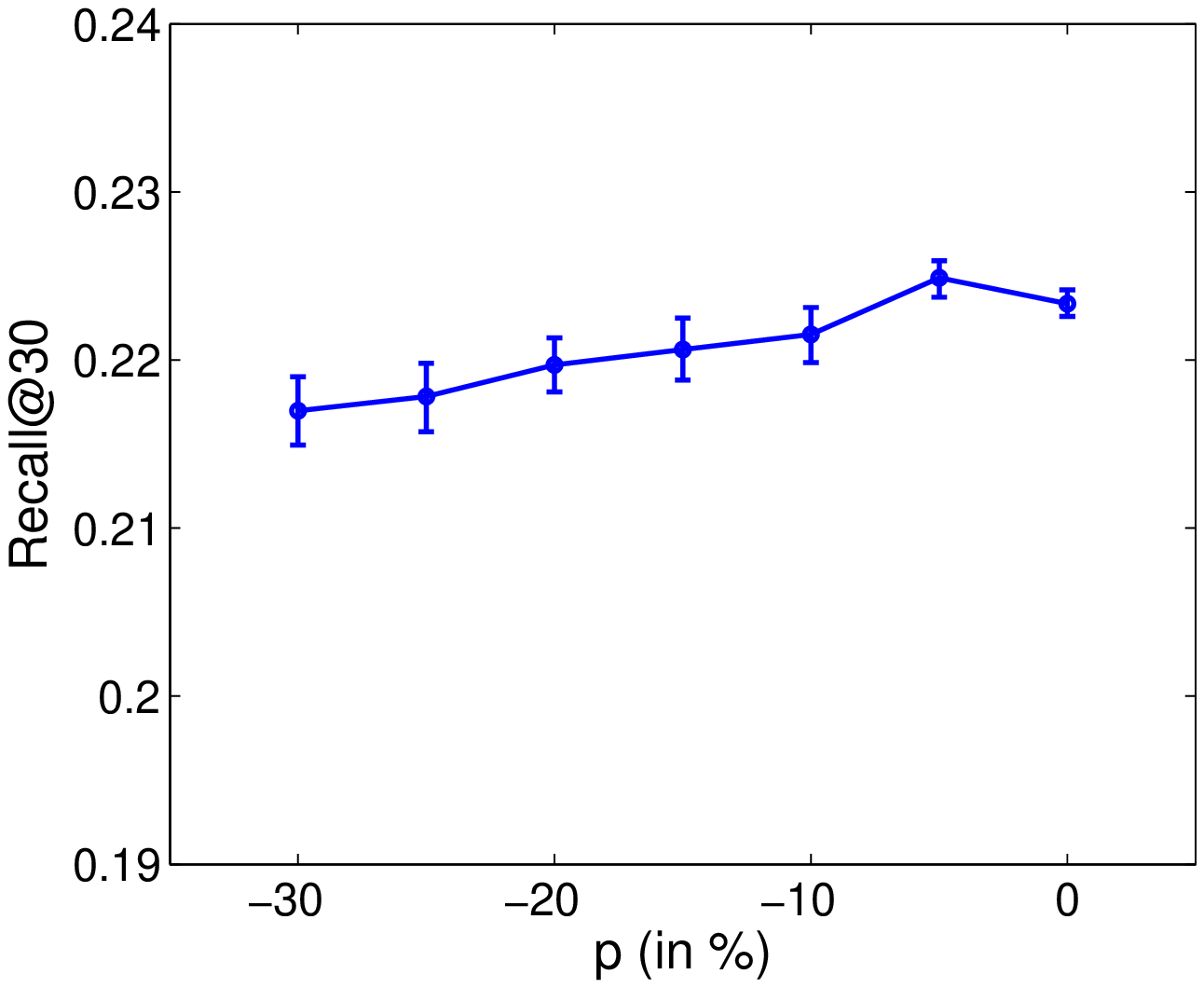}}
  \caption{(a) Recall@30 for different levels of random friend addition/deletion noise with 95\% confidence intervals.  (b) Recall@30 for different levels of random user deletion with 95\% confidence intervals.}
  \label{recallDiffFriendPlot}
  \end{center}
\end{figure}

We next compare the SCR approach with other state-of-the-art algorithms.  Recall that, for the Last.fm dataset described in the previous section, a $\text{\emph{query}}\times\text{\emph{user}}\times\text{\emph{item}}$ tensor entry corresponds to a $\text{\emph{genre}}\times\text{\emph{user}}\times\text{\emph{artist}}$ triple, where genres are obtained from the set of filtered user tags.  We preprocess the data set in two ways to obtain listening counts for each such triple/tensor entry.  First, if a user $u$ has listened to an artist $a$ and assigned multiple genres, for example \emph{rock} and \emph{indie}, then $u$'s listening counts for $a$ are evenly distributed to triples $(\emph{rock},u,a)$ and $(\emph{indie}, u, a)$. If the user has not assigned any genre to an artist, the genres of $a$ are those assigned to $a$ by other users, and the listening count of $a$ is distributed to each triple according to how frequently the genre appears. If no user has ever assigned any genre to $a$, the genres of $a$ are defined as the genres used by $u$ for any artist, and the listening count is again prorated to the triples. 

Second, since we are interested in ranking artists given a particular user and query, we normalize listening counts of triples having the same $u$ and $q$ so that their weights sum to 1. In the end we have 389,405 data points of the form $(q,u,a,w)$, where $w$ is the normalized listen count of artist $a$ by user $u$ in genre $q$.  

Since our main goal is to show how social information can help compensate for data sparsity in a collaborative retrieval task, we identify a series of subsets of the Last.fm data that correspond to increasingly less compact social networks.  We use a standard implementation of hierarchical clustering on the complete social adjacency matrix to select subsets of users that exhibit significant internal social structure; the number of users in these sets varies from 200 to 1000 (see Table~\ref{dataset}).  For each user set, the corresponding set of items contains all artists listened to by one or more of the selected users.  In this way, the number of artists grows organically with the number of users.  As in Section~\ref{SIA}, we use the 30 most frequent genre tags as our query set. 

The resulting datasets are referred to as Compact-\emph{lastfm-N},
where $N$ denotes the number of users in the dataset.  Their statistics are shown in Table~\ref{dataset}.  By construction, users in the smaller datasets have higher average numbers of within-set friends.  This means that the smaller sets are more tightly connected, which may make them more amenable to social regularization.  Conversely, the larger datasets are sparser and may be more representative of large-scale social networks.  Note that the \textit{density} of social links falls with the number of users even if the average number of within-set friends stays constant, thus the largest datasets are in fact quite a bit more sparse (relatively speaking) than the smallest ones.  We will show how the performance of SCR changes as these qualities are varied.

\begin{table*}[t]
\begin{center}
\caption{Last.fm dataset statistics}
\begin{tabular}{|c|c|c|c|c|c|c|}
  \hline   
   Dataset &  users & items (artists) & queries (tags) & samples & data sparsity (\%) & average \# of friends \\
  \hline \hline 
  \emph{lastfm-2k} & 1892 & 17632 & 11946 & 186479 & 99.9999 & 13.44 \\
   \hline
   Compact-\emph{lastfm-200} & 200 & 2392 & 30 & 29850 & 99.7920& 28.54 \\
   \hline
   Compact-\emph{lastfm-300} & 300 & 3299 & 30 & 44318 & 99.8507  & 29.79 \\
   \hline
   Compact-\emph{lastfm-400} & 400 & 4091 & 30 & 58098 & 99.8817 & 29.10 \\
   \hline
   Compact-\emph{lastfm-500} & 500 & 4928 & 30 & 72125 & 99.9024 & 27.81 \\
   \hline
   Compact-\emph{lastfm-600} & 600 & 5765 & 30 & 85522 & 99.9176 & 26.32 \\
   \hline
   Compact-\emph{lastfm-700} & 700 & 6454 & 30 & 98367 & 99.9274 & 24.90 \\
   \hline
   Compact-\emph{lastfm-800} & 800 & 7071 & 30 & 111062 & 99.9346  &  23.57 \\
   \hline
   Compact-\emph{lastfm-900} & 900 & 7782 & 30 & 124005 & 99.9410 & 22.23 \\
   \hline
   Compact-\emph{lastfm-1000} & 1000 & 8431 & 30 & 137518 & 99.9456 & 21.01 \\
   \hline
  
\end{tabular}
    \label{dataset}
\end{center}
\end{table*}

\subsection{Evaluation}

We compare SCR with state-of-the-art algorithms used for collaborative retrieval as well as traditional collaborative filtering.  Popular matrix factorization methods such as SVD and NMF are often used for collaborative filtering; these methods optimize the deviation of the rating matrix from entries supplied in the training set.  However, standard SVD and NMF techniques are not directly applicable to tensors.  Instead, we perform NMF on the $|\mathcal{Q}|$ different user $\times$ artist matrices to compute the rank of $a$ among all artists given $q$ and $u$. We also compare to latent collaborative retrieval (LCR). 

The dimension of the embeddings for all methods is chosen to be 30; as shown in Figure~\ref{recallDiffDimensionPlot}, this choice yields approximately optimal performance for SCR-generalized; however, the results are not qualitatively different for other choices of embedding dimension.  The hyperparameters $w_s$, $\eta$, and $c$, along with constraint parameters $L_S$, $L_T$ and $L_V$, are chosen separately for each method (as applicable) using a validation set (see below). Since matrix factorization approaches are not specially designed for tensors and typically show worse performance than LCR \citep{weston2012latent}, we only present results for NMF, which performed the best.  We use the NMF implementation from \url{http://www.csie.ntu.edu.tw/~cjlin/nmf/}.  For each experiment, 60\% of the samples are used for training (or less; see below), 20\% are used for validation, and 20\% are used for testing.

To evaluate the performance of each algorithm, for a given test sample $(q,u,a,w)$ we first compute $f(q,u,i)$ for $i = 1,\dots,|\mathcal{A}|$ and sort the artists in descending order of score.  We then measure recall@$k$, which is 1 if artist $a$ appears in the top $k$, and 0 otherwise, and report the mean \emph{recall}@$k$ over the whole test dataset.  As a secondary measure we report \emph{weighted recall}@$k$, which is the relevance score $w$ if artist $a$ appears in the top $k$, and 0 otherwise.  Mean weighted recall@$k$ thus not only measures how many triples are ranked in the top $k$, but the quality of these test triples.

\subsection{Results}

We begin with results for the smallest datasat, Compact-\textit{lastfm-200}, which is small enough to be practical for all methods.  The resulting recall@$k$ and weighted recall@$k$ for different $k$ are shown in Figure \ref{recallPlot}.  SCR (generalized or standard) outperforms the baselines on this top $k$ ranking problem; note that SCR-generalized outperforms SCR under the weighted recall criterion, which makes sense since it incorporates relevance scores in the loss function.

Since we expect social information to be particularly useful when data are sparse, we also show results of recall@$k$ and weighted recall@$k$ for different amounts of training data $(100\%, 80\%, 60 \% \text{ and } 40\% \text{ of the total training data})$ in Figure \ref{recallDiffTrainPlot}. Notice that the performance gap between SCR-generalized and LCR becomes larger as the number of available training examples is reduced, suggesting that our proposed algorithm can be especially useful for predicting the interests of new users or infrequent users when social network information is available.

Moving to the larger datasets, computation time increasingly becomes an issue for the matrix factorization approaches, so we focus on only two algorithms: SCR-generalized and LCR.  Figure~\ref{recallDiffDatasetPlot} shows recall results for all of the compact datasets.  Note that the performance gap between SCR and LCR narrows slightly but remains significant even as the size of system becomes larger and the density of social links decreases.  It may be counterintuitive that performance decreases (at least for unweighted recall) as the size of the dataset grows; however, since the number of artists grows with the number of users, the prediction problem is becoming more difficult at the same time.
These results suggest that, while a dense social network may improve the relative performnce of SCR, it retains significant advantages even in larger, sparser settings.

Finally, we show in Figure~\ref{timeComparison} the runtimes for each stochastic gradient training iteration of SCR and LCR on the Compact-\textit{lastfm-200} dataset; SCR is dramatically faster, despite using essentially similar optimization techniques.  This is because the runtime is dominated by the sampling procedure used to estimate the rank function.  LCR promotes the observed items to high positions quickly, thus making subsequent iterations quite slow.  SCR, on the other hand, has additional regularization that appears to prevent this situation.  Combined with the performance improvements discussed above, this is a significant practical advantage.

\begin{table}[t]
  \centering
  \caption{Average number of friends added/removed per user.}
  \begin{tabular}{|cc|cc|}
    \hline
    $p$ & Friends removed & $p$ & Friends added\\
    \hline
    \hline
        -0.03 & 1.59              & 0 & 0          \\
        -0.06 & 3.27              & 0.03 & 1.62  \\
        -0.09 & 4.84              & 0.06 & 3.35  \\
        -0.12 & 6.27              & 0.09 & 5.04  \\
        -0.18 & 9.10              & 0.12 & 6.68  \\
        -0.24 & 11.71             & 0.18 & 10.00  \\
        -0.30 & 14.25             & 0.30 & 16.67 \\
        -0.36 & 16.33             &&               \\
        -0.42 & 18.38             &&               \\
        -0.48 & 20.18             &&               \\
        -0.60 & 23.25             &&               \\
    \hline
  \end{tabular}
  \label{ptable}
\end{table}

\begin{figure*}[t]
\begin{center}
\subfloat[]{
  \includegraphics[trim=0cm 5.5cm 0cm 5.5cm, clip,width=5.5cm]{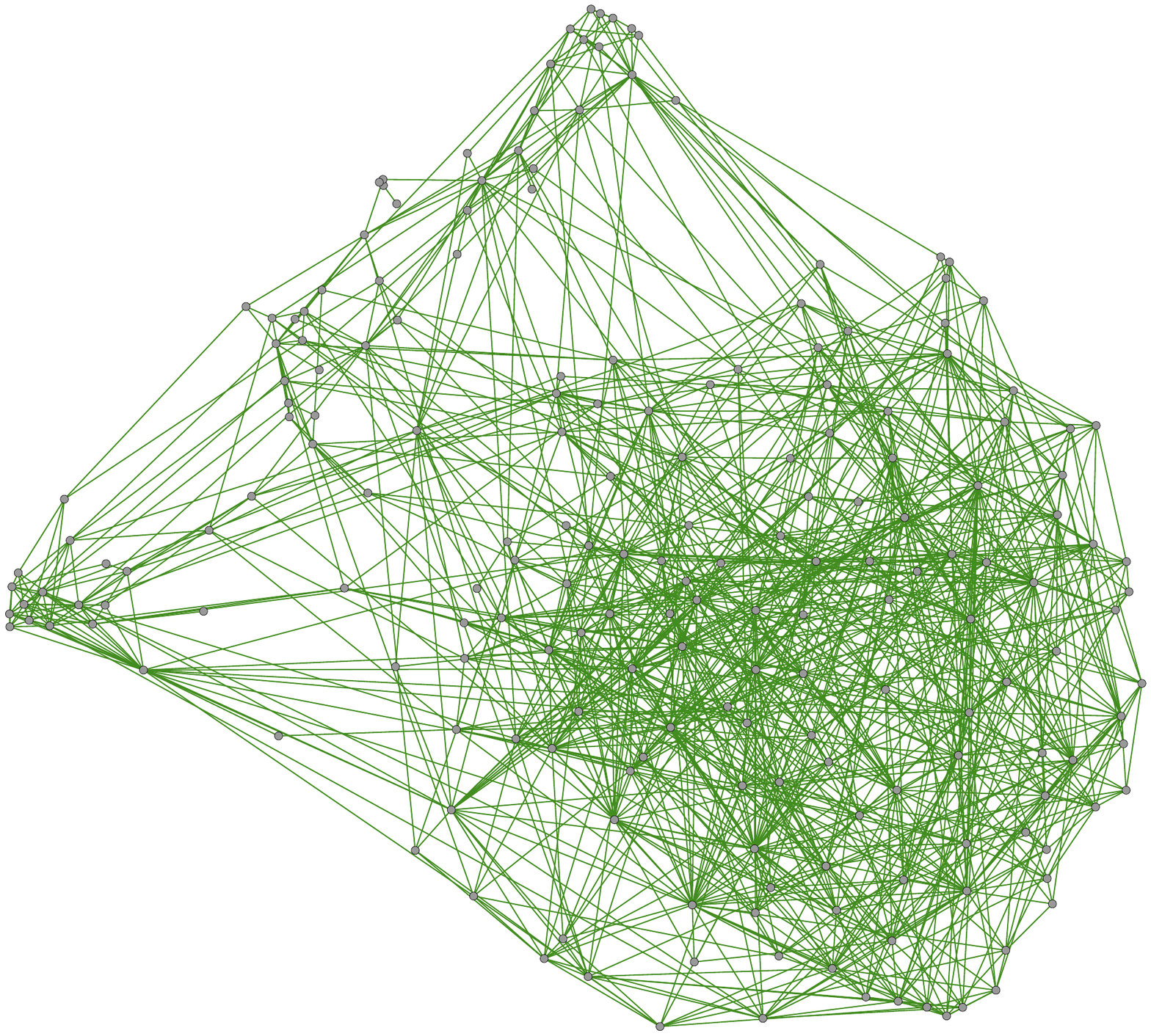}}
\hspace{2cm}
  \subfloat[]{
  \includegraphics[trim=0cm 5.5cm 0cm 5.5cm, clip,width=5.5cm]{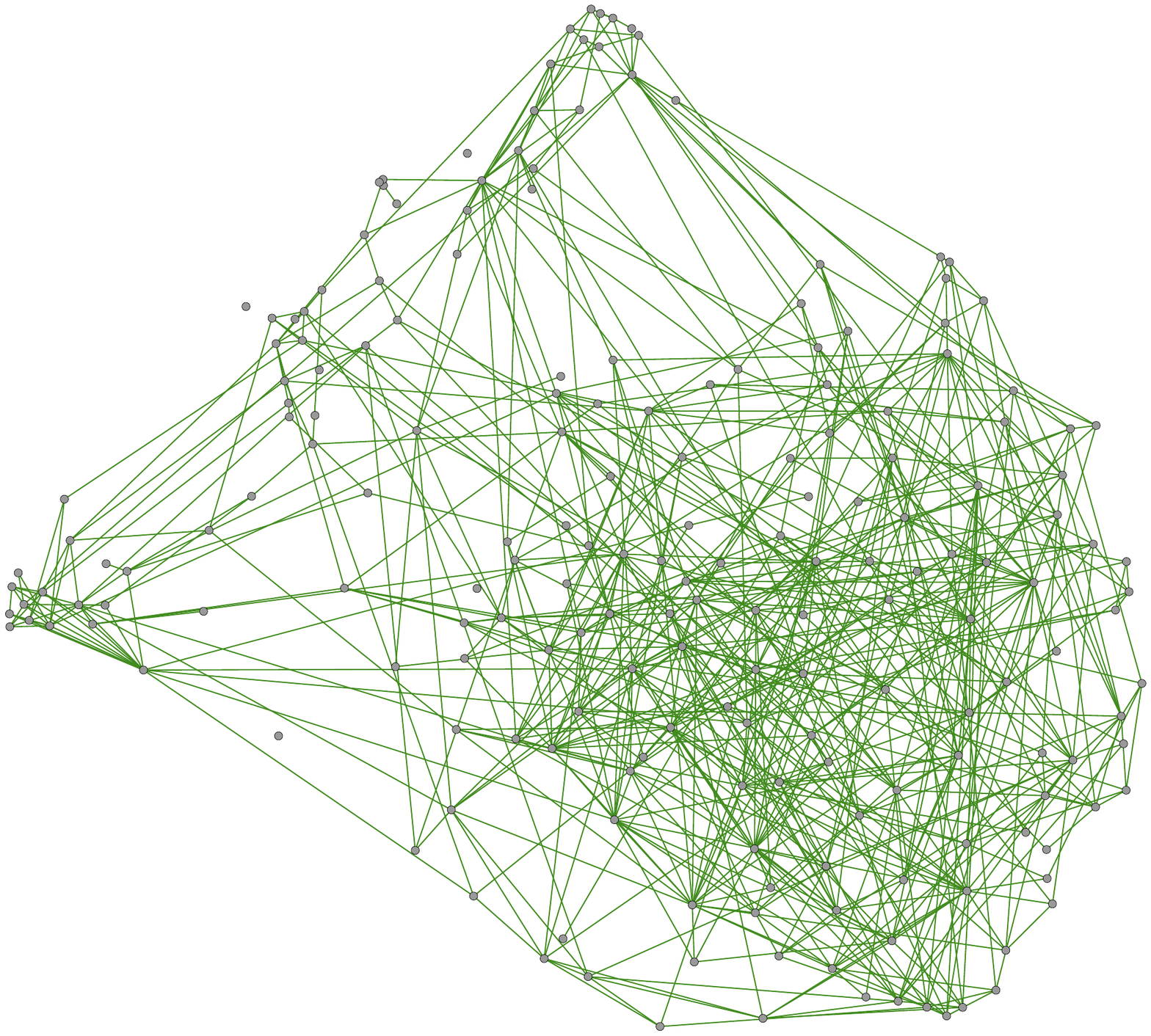}}\\
   \subfloat[]{
  \includegraphics[trim=0cm 5.5cm 0cm 5.5cm, clip,width=5.5cm]{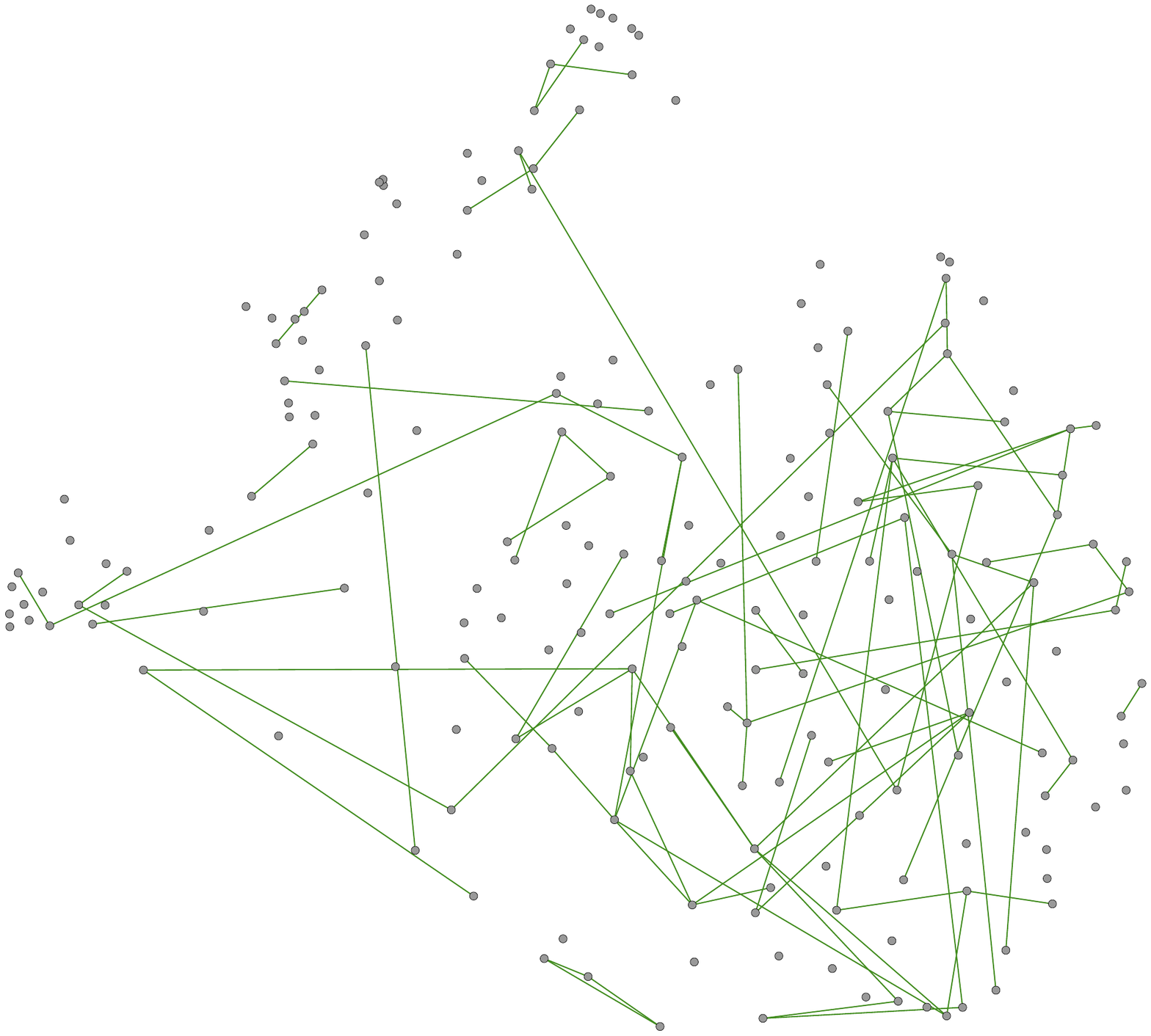}}
\hspace{2cm}
\subfloat[]{
  \includegraphics[trim=0cm 5.5cm 0cm 5.5cm, clip,width=5.5cm]{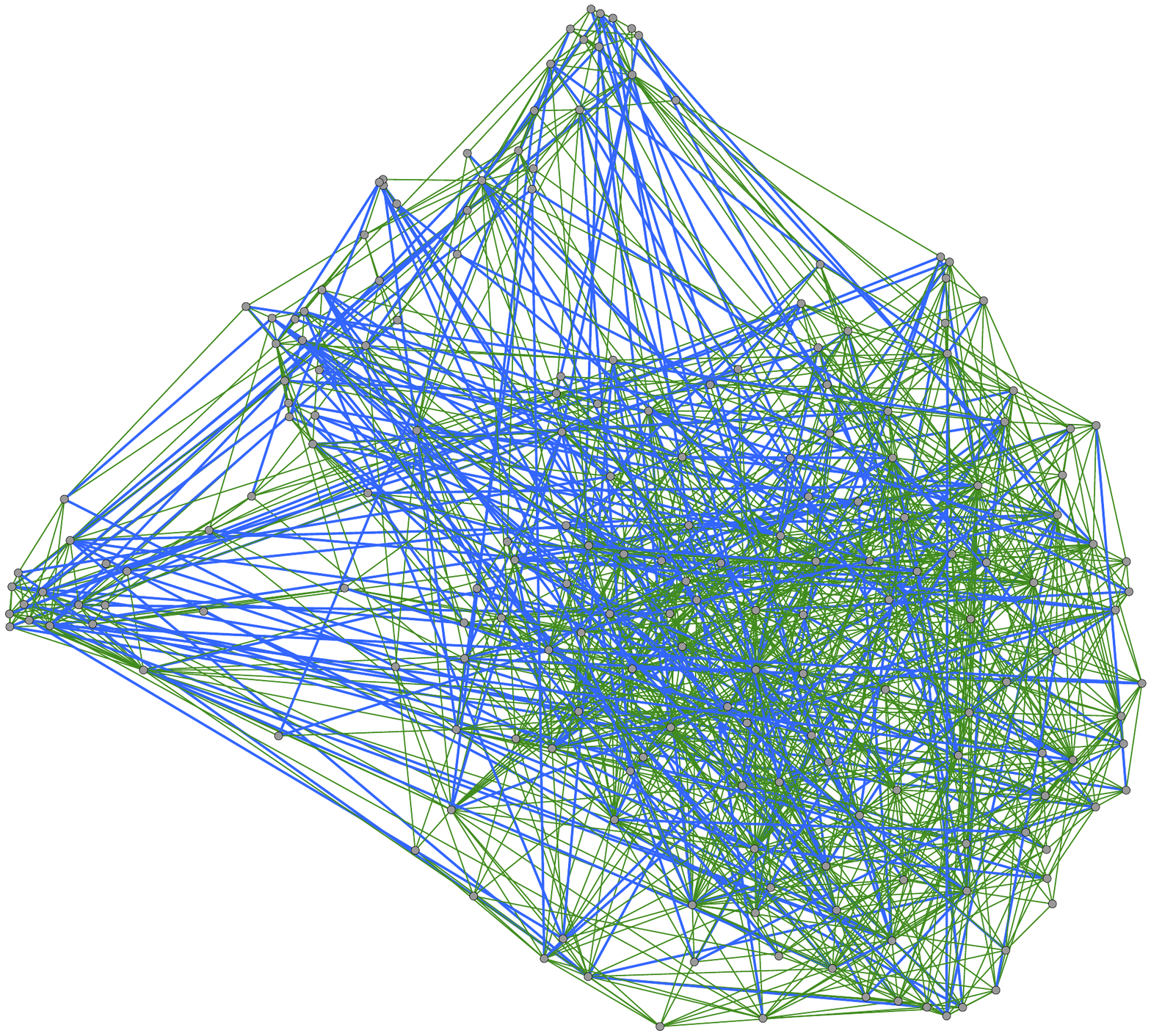}}
  \caption{ Visualizations of 200-user social graphs with different levels of random friend addition/deletion noise. (a) Original social graph ($p = 0$). (b) Reduced social graph ($p = -0.30$). (c) Reduced social graph ($p = -0.60$). (d) Augmented social graph ($p = 0.09$), with added edges shown in blue. The arrangement of the vertices was generated using the ForceAtlas2 algorithm \cite{jacomy2011forceatlas2}. } 
  \label{graphVisualization}
  \end{center}
\end{figure*}

\subsection{Sensitivity}

Since real-world social networks are subject to various sources of noise, we test the robustness of SCR on a series of datasets in which edges in the social graph have been randomly added or removed.  Specifically, in these experiments we begin with the Compact-\textit{lastfm-500} dataset, and then, for a user with $F$ friends, randomly add $Fp$ friend relations for a noise parameter $p$.  If $p$ is negative, then we instead remove $F|p|$ edges from the original graph.  Figure~\ref{recallDiffFriendPlot}(a) shows the results of SCR learning using these noisy datasets for various values of $p$, averaged over 25 random trials.  Table~\ref{ptable} shows the average number of friends added or removed per user for a given $p$.

The results reveal an interesting, asymmetric behavior.  When friends
are added at random ($p > 0$), performance begins to drop quickly, presumably due to the fact that non-friends can have significantly different preferences, as shown in Section~\ref{SIA}.  Luckily, the creation of spurious non-friend edges in the social graph seems relatively unlikely in the real world, where links must typically be confirmed by both parties.

On the other hand, removing edges ($p < 0$) seems to have a relatively small impact on performance unless a significant proportion of the links are removed.  This may be because friends are often linked by multiple short paths though other mutual friends, thus the removal of a single link only slightly diminishes connectivity.  Moreover, groups of users linked in cliques tend to influence each other strongly, and such cliques cannot be broken up by removing only a few edges.  This type of noise, though presumably common, has only a limited impact when using social networks for collaborative retrieval.

To visualize these patterns, we select a random subset of 200 users and plot the original social graph as well as the social graphs obtained for different values of $p$ in Figure \ref{graphVisualization}.  When $p=-0.3$, the basic structure of the network is still visible, and almost every user is still connected to all of his or her original friends through short paths in the reduced graph. However, when $p= -0.6$, the structure has begun to break down, and many former friends are now totally disconnected.  In this case the performance of SCR degrades approximately to the level of LCR.  When $p$ is positive, we see a different kind of degredation, where many new edges connect together subgroups that were originally only sparsely connected.

An alternative test of sensitivity is to simulate incomplete membership by randomly deleting users from the network. We begin with the same Compact-\emph{lastfm}-500 dataset and randomly delete $500|p|$ users from the network. Figure~\ref{recallDiffFriendPlot}(b) shows the results for various values of $p$, averaged over 25 random trials. It shows that when random users are deleted from the network, there is, on average, a noticeable but relatively minor decrease in recall. This is what we would expect, since fewer users means there is less social information to leverage, as well as less training data.

\FloatBarrier
\section{Conclusions and future work}
We proposed SCR, a new model blending social networking, information retrieval, and collaborative filtering. SCR uses a social graph to improve performance on collaborative retrieval problems, which we believe are increasingly important in practice, outperforming state-of-the-art CR and CF approaches.  We also showed that users tend to share interests with friends.  Going forward we hope to develop a two-pass version of the SCR algorithm that helps predict interest commonalities between friends, and can be used to prune out edges on the social graph that may work against achieving good performance. 

\bibliographystyle{IEEEtranN}
\bibliography{ReferenceSCR.bib} 

\end{document}